\documentstyle[12pt,a4]{article}

\newcommand{\re}[1]{(\ref{#1})}
\newcommand{\be}{\begin{equation}}
\newcommand{\ee}{\end{equation}}
\newcommand{\bea}{\begin{eqnarray}}
\newcommand{\eea}{\end{eqnarray}}
\newcommand{\beas}{\begin{eqnarray*}}
\newcommand{\eeas}{\end{eqnarray*}}
\newcommand{\no}{\nonumber}
\newcommand{\n}{{\vec{n}}}
\newcommand{\m}{{\vec{m}}}

\renewcommand{\i}{{\vec{i}}}
\newcommand{\k}{{\vec{k}}}
\newcommand{\Th}{{\bf {\rm Th}}}

\newcommand{\ab}[1]{{\bf #1}}
\renewcommand{\S}{{\cal S}}
\newcommand{\T}{{\cal T}}

\title{Lattice electrons in constant magnetic field:
	Bethe like ansatz}

\author{T.Hakobyan\\
	\small{Yerevan Physics Institute}\\
	\small{375036 Alikhanian Br.2, Yerevan, Armenia}\\
	\footnotesize{e-mail: hakob@vx1.erphy.armenia.su}
	\\ \\
	A.Sedrakyan\\
	\small{Yerevan Physics Institute}\\
	\small{375036 Alikhanian Br.2, Yerevan, Armenia}\\
	\footnotesize{e-mail: sedrak@vx1.erphy.armenia.su}\\
	}

\vspace{2cm}

\date{October 1994}

\begin{document}

\begin{titlepage}

\maketitle

\begin{abstract}

We
use the functional representation  of Heisenberg-Weyl  group and
obtain equation for the spectrum of the model,  which is  more complicated
than Bethes ones, but can  be written explicitly through theta
functions.

\end{abstract}

\vspace{3cm}

Preprint {\bf YERPHI-1427(14)-94}

\end{titlepage}

\newpage

\noindent

The Azbel-Hofstadter problem, i.e. the problem of Bloch electrons in magnetic
field \cite{Zak,Azbel,Wannier,Hofstadter} recently attracts the attention of
both condensed matter and string physicists
\cite{Wiegmann1,Wiegmann2,Callan,Faddeev}.
It had been shown that dissipative Hofstadter model exhibits critical behavior
on a network of lines in the dissipation magnetic field plane and the
corresponding critical theories represent nontrivial solution of open
string field theory, as it was observed by Callan,Folce and Freed in
\cite{Callan}. From the other side Wiegmann and
Zabrodin \cite{Wiegmann1,Wiegmann2} by making explicit natural relation between
the
group of magnetic translations and quantum group $U_qsl_2$, using its
functional representation, obtained Bethe Ansatz equations for energy
spectrum of homogeneous model for one particular value of quasimomentum
(midband point)

 In \cite{Faddeev} Faddeev  and Kashaev  obtained some
Bethe Ansatz equations for all values of momentum by using another
method. The functions in these equations are defined
on  some higher genus algebraic  curve but the explicit solution is obtained
only for the same  particular value of quasimomentum
in the nonhomogeneous case.

In   this  paper instead of reduction of the problem to $U_qsl_2$ we
use the functional representation  of Heisenberg-Weyl  group and
obtain equation for the spectrum of the model,  which is  more complicated
than Bethes ones, but can  be written explicitly through theta
functions. Considered representation lead to cyclic representation
of $U_qsl_2$ via Heisenberg-Weyl algebra, introduced in \cite{Floratos}.

\vspace{1cm}

Let  us recall the contents of the electron model in a constant magnetic
field.

\indent
The Hamiltonian is given by
\bea
\label{Hamil}
H=\sum_{<\n,\m>} t_{\n,\m}e^{iA_{\n,\m}}c_\n^+c_\m
\eea

Here  $c_\n,c_\n^+$ are the annihilation and creation operators of electron
at site $\n$  of a
two dimensional rectangular lattice, $A_{\n,\m}=-A_{\m,\n}$ is the vector
potential
of a  magnetic  field with the  strength perpendicular to the plane of
lattice  and $t_{\n,\m}$  are hoping   amplitudes between the neighbors.
We choose  them  $t_x$ and  $t_y$ in the horizontal and vertical directions
correspondingly. For the homogeneous model $t_x=t_y$.

Note that the hamiltonian \re{Hamil} commutes with the total particle
number operator $N=\sum_{\n}c_\n^+c_\n$. In the following we consider
the diagonalization problem of \re{Hamil} in one-particle sector.
In this sector the action of translation operator by vector $\vec{\mu}$
can be written by
$$
S_{\vec{\mu}}=\sum_\i\vert\i\rangle\langle\i+\vec{\mu}\vert,
$$
where the standard bra- and ket- vectors are used:
$\vert\i \rangle=c_\i^+|vac\rangle$, $\langle\i\vert=\langle vac |c_\i$.

We consider  the case when the magnetic  flux per plaquette
\begin{equation}
\label{plaq}
\exp(i\Phi)=\prod_{plaquette} \exp(iA_{\n,\m})
\end{equation}
 is rational:
$\Phi=2\pi {P\over Q}$, where $P$ and $Q$ are mutually prime  integers.
 The product in  \re{plaq} is performed on anticlockwise
direction.

\indent
Taking different  gauges one  can obtain various equivalent forms of
\re{Hamil}. In the following we will use the Landau gauge, which  is
 defined by
$$
 A_x=A_{\n,\n+\vec{1}_x}=0 \ \ A_y=A_{\n,\n+\vec{1}_y}=\Phi n_x
$$

Then the Hamiltonian (\ref{Hamil}) is invariant with respect to
the translations $S_{\pm y}=S_{\pm \vec{1}_y}$ and
$S_{\pm x}^Q=S_{\pm\vec{1}_x}^Q$:
$$
{[}H,S_{\pm x}^Q{]}={[}H,S_{\pm y}{]}=0
$$
So, the problem  of diagonalization of $H$ reduced to its diagonalization
on each  eigenspace of $S_x^Q$ and $S_y$ (\cite{Hofstadter}).
The latter is a $Q$-dimensional  space  $\Psi(\k)$, spanned by
Bloch wave functions
$$
\psi_{n_x}(\k) = \sum_{i_x=n_x {\rm mod}Q \atop i_y} e^{-i \k\i}
 \vert\i\rangle,
$$
which   satisfy
\bea
\label{trrep}
& \psi_{n_x}(\k) =\psi_{n_x+Q}(\k) &
\no\\
& S_{\pm y}\psi_{n_x}(\k)=e^{\mp ik_y}\psi_{n_x}(\k) &\\
& S_{\pm x}\psi_{n_x}(\k)=e^{\mp ik_x}\psi_{n_x\mp1}(\k) & \no
\eea

Let us define also the generators  of magnetic translations by
$$
T_{\vec{\mu}}=\sum_\i e^{iA_{\i,\i+\vec{\mu}}}\vert\i\rangle
  \langle\i+\vec{\mu}\vert
$$
for $\vec{\mu}=\vec{\pm1_x},\vec{\pm1_y}$.
They satisfy the following  relations

\bea
\label{Heisenberg}
T_{\pm y}T_{\pm x}=q^{-2}T_{\pm x}T_{\pm y} &
T_{\pm y}T_{\mp x}=q^{2}T_{\mp x}T_{\pm y}
\eea
and  form   Heisenberg-Weyl group. Here we used the notation
$$
q=\exp\left(i\frac{\Phi}{2}\right)=\exp\left(\pi i\frac{P}{Q}\right)
$$

Note, that  is any gauge $S_{x,y}^Q=T_{x,y}^Q$
(in the Landau gauge moreover $S_x=T_x$).

The action of magnetic translations $T_{\pm x}$, $T_{\pm y}$
  on Bloch functions $\psi_{n_x}(\k)$ has the following form:
\bea
\label{magrep}
&T_{\pm y}\psi_{n_x}(\k)=e^{\mp ik_y\pm in_x\Phi}\psi_{n_x}(\k) &   \no\\
& T_{\pm x}\psi_{n_x}(\k)=e^{\mp ik_x}\psi_{n_x\mp 1}(\k) &
\eea

\indent

The Hamiltonian \re{Hamil} can be written in terms of  the generators of
magnetic translations by  $H=t_x(T_x+T_{-x})+t_y(T_y+T_{-y})$
(\cite{Wiegmann1,Wiegmann2}).

Note that on $\Psi(\k)$  the $Q$-th power  of magnetic translations
are scalars:
\beas
\label{center}
& T_{\pm x}^Q =e^{\mp iQk_x}\cdot {\rm id} \qquad T_{\pm y}^Q = e^{\mp iQk_y}
\cdot {\rm id}&
\eeas

\vspace{1cm}

To define the functional form of equation $H\psi=E\psi$ let us recall
the  representation of Heisenberg-Weyl group \re{Heisenberg} on  the space of
complex functions (\cite{Manford}).
It can be constructed in the following way.

Define the  actions $\S_b$ and $\T_a$, $a,b\in Z$
on the space of analytic functions on the complex  plain  as
\bea
\S_bf(z)=f(z+b) & \T_a(\tau)f(z)=\exp(\pi ia^2\tau+2\pi iaz) f(z+a\tau)
\no
\eea
for some $\tau\in C$. Then
\bea
\S_a\S_b=\S_{a+b} & \T_a(\tau)\T_b(\tau)=\T_{a+b}(\tau) & \S_a\T_b(\tau)
	=\exp(2\pi iab)\T_b(\tau)\S_a
\no
\eea

Consider the space of theta functions with characteristics
$\Theta^{(Q)}(z,\tau)$,
which are invariant with respect to subgroup, generated by
$\S_{\pm Q}$ and $\T_{\pm 1}(\tau)$. They form the Q-dimensional space
$\Th_Q(\tau)$
and have in the fundamental domain with vertices $(0,\tau,Q,Q+\tau)$
precisely $Q$ zeroes.

The space $\Th_Q(\tau)$
forms an irreducible representation of Heisenberg-Weyl group \re{Heisenberg},
generated by
\bea
\label{T_x}
& T_{\pm x}:=\alpha_\pm \T_{\mp {1\over Q}}(\tau) \qquad
 	T_{\pm y}=\beta_\pm \S_{\pm{P}}=\beta_\pm \S_{\pm{1}}^P &
\eea
One can choose the basis $\Theta_{a,0}(z,\tau)$ of $\Th_Q$ as follows:
$$
\Theta_{a,0}(z,\tau)=(\T_{a}(\tau)\Theta)(z,\tau),
$$
where $ a= 0,1/Q,\dots,(Q-1)/Q$.
and $\Theta(z,\tau)=\Theta_0(z,\tau)$ is standard theta function.
\footnote{We used the standard notation of theta functions with
characteristics $\left[\begin{array}{c}
N\\
M
\end{array}\right]$:
$$\Theta_{a,b}(z,\tau)=\Theta\left[\begin{array}{c}
a\\
b
\end{array}\right](z,\tau)=T_aS_b\Theta(z,\tau)$$,
where $a\in {1\over N}\ab{Z} \
{\rm mod 1}$, $b\in {1\over M}\ab{Z} \ {\rm mod 1}$
(\cite{Manford}).}
In this basis
\beas
\S_{\pm 1}\Theta_{a,0}=\exp(\pm {2\pi ia})\Theta_{a,0}
\no\\
\T_{\pm{1\over Q}}\Theta_{a,0}=\Theta_{a\pm {1\over Q},0}
\eeas
Comparing these equations with \re{magrep},\re{center}
we obtain for parameters $\alpha_\pm,\, \beta_\pm$
\bea
&\alpha_\pm=\exp(\mp ik_x) & \beta_\pm =\exp(\mp ik_y) \no
\eea
The space $\Psi(\k)$ is identified with the space of theta
functions with characteristic $\left[\begin{array}{c}
Q\\
1
\end{array}\right]$:
$$
\psi_{n_x}(\vec{k})\sim \Theta_{{n_x\over Q},0}(z,\tau)
$$

The representation \re{magrep}, \re{T_x}, of
course, is equivalent to standard one by $Q\times Q$ matrices:

$$
T_{x}=\alpha_+\left(\begin{array}{lllll}
  0 & 1 &  0 &  \dots  & 0 \\
 \dots &  \dots & \dots &\dots  &\dots  \\
  0 &  0 & 0 &  \dots & 1 \\
  1 &  0 & \dots& 0 & 0
\end{array} \right)
$$

$$
T_{ y}=\beta_+\left(\begin{array}{ccccc}
  1 & 0 &  0 &  \dots  & 0 \\
  0 &  q^{2} & 0 &  \dots & 0 \\
 \dots &  \dots & \dots &\dots  &\dots  \\
  0 &  0 & 0 &  \dots & q^{2(Q-1)}
\end{array} \right)
$$

Note that one can represent the generators $J_{\pm},J_0$ of quantum group
$U_qsl_2$
\beas
&[J_+,J_-]=\frac{q^{2J_0}-q^{-2J_0}}{q-q^{-1}} \qquad
	q^{J_0}J_\pm q^{-J_0}=q^{\pm1}J_\pm&
\eeas
in terms of magnetic translations (\cite{Wiegmann1,Wiegmann2}):
\bea
\label{QGrep}
&J_+=\frac{t_xT_{ x}+t_yT_{y}}{q-q^{-1}}  \qquad
J_-=\mp\frac{t_xT_{ -x}+t_yT_{-y}}{q-q^{-1}}&\no\\
&q^{2J_0}=\pm qT_xT_{-y}=\pm q^{-1}T_{-y}T_x &
\eea
This is the cyclic representation of $U_qsl_2$, earlier constructed by
Floratos in \cite{Floratos}.

The functional representation of $U_qsl_2$ for spin-$j$ representations,
used in \cite{Wiegmann1,Wiegmann2} for solving \re{Hamil} for special value of
momentum by means of Bethe Ansatz, has also the structure of \re{QGrep}, where
$t_{x,y}$ and the generators of Heisenberg group are defined by the following
\bea
\label{Wfunc}
&T_{\pm x}\psi(z)=-q^{\mp j+{1\over 2}}z^{\pm1}\psi(q^{\pm1}z) &
T_{\pm y}\psi(z)=q^{\pm j-{1\over 2}}z^{\pm1}\psi(q^{\mp1}z) \no\\
&t_x=q^{-j-{1\over 2}} & t_y=q^{j+{1\over 2}}
\eea
Note that the representation \re{Wfunc} of $T_{\pm x},T_{\pm y}$  is
infinite dimensional in contrast with the representation \re{T_x}.
It becomes finite dimensional after its restriction \re{QGrep} on quantum
group.

\vspace{1cm}

The space $\Th_Q(\tau)$ have some analog with the space of polynomials
of degree $Q-1$. The polynomial decomposition analog for the $\Theta^{(Q)}$
is the following.

Every $\Theta^{(Q)} \in \Th_Q$ can be represented in the
form
\beas
&&\Theta^{(Q)}(z,\tau)=\\
&&\quad\gamma\exp(-2\pi ikz/Q)\Theta\left(\frac{z-z_1}Q,\frac\tau Q\right)
	\Theta\left(\frac{z-z_2}Q,\frac\tau Q\right)
        \ldots \Theta\left(\frac{z-z_Q}Q,\frac\tau Q\right),
\eeas
where
$$
\sum_{i=1}^Q z_i=k\tau+nQ, \ \ k,n\in\ab{Z}, \ \ 0\le k,n <Q, \ \
\gamma\in \ab{C}
$$
Then for $H=T_x+T_{-x}+T_y+T_{-y}$ the Schr\"odinger equation
 $H\psi=E\psi$ after rescaling of the arguments of $\Theta(z,\tau)$:
$z\to Qz$, $\tau\to Q\tau$ can be written in terms
of products of theta functions. We'll write them in terms of
$$
\Theta_1(z,\tau):=\exp\left({\pi i\tau \over 4}+{\pi i\over 2}+
\pi iz\right) \Theta\left(z+{\tau+1\over 2},\tau\right),
$$
which obey $\Theta_1(0,\tau)=0$,
by performing additional substitution: $z\to z+{{1+\tau}\over 2}$.

\bea
(-1)^P\beta_+ \exp\left(-\frac{2\pi ikP}{Q}\right)
	\prod_{i=1}^Q \Theta_1(z-z_i+{P\over  Q},\tau) &+&\nonumber\\
  (-1)^P\beta_- \exp\left(\frac{2\pi ikP}{Q}\right)\prod_{i=1}^Q
	\Theta_1(z-z_i-{P\over  Q},\tau) &- &
\no\\
\alpha_+ \exp\left(\pi i\tau\frac{1+2k}Q-2\pi iz\right)
        \prod_{i=1}^Q \Theta_1(z-z_i-{\tau\over  Q},\tau) &-&
\label{eq:B}\\
  \alpha_- \exp\left(\pi i\tau\frac{1-2k}Q+2\pi iz
	\right)
        \prod_{i=1}^Q \Theta_1(z-z_i+{\tau\over  Q},\tau) &=&
\no\\
E\prod_{i=1}^Q \Theta_1(z-z_i,\tau)&&
\no
\eea

 The roots of right and left side of (\ref{eq:B}) must
coincide. So, inserting the zeroes of $\Theta_1$ $z=z_i, \  i=1,\dots,Q$, we
obtain the system of $Q$ equations for the parameters $z_i$:

\bea
(-1)^P\beta_+ \exp\left(-\frac{2\pi ikP}{Q}\right)
	\prod_{i=1}^Q \Theta_1(z_j-z_i+{P\over  Q},\tau) &+&
\no\\
(-1)^P \beta_- \exp\left(\frac{2\pi ikP}{Q}\right)
	\prod_{i=1}^Q \Theta_1(z_j-z_i-{P\over  Q},\tau) &-&
\label{eq:Bt}\\
\alpha_+ \exp\left(\pi i\tau\frac{1+2k}Q-2\pi iz_j\right)
        \prod_{i=1}^Q \Theta_1(z_j-z_i-{\tau\over  Q},\tau) &-&
\no\\
  \alpha_- \exp\left(\pi i\tau\frac{1-2k}Q+2\pi iz_j\right)
        \prod_{i=1}^Q \Theta_1(z_j-z_i+{\tau\over  Q},\tau) &=& 0
\no
\eea
This is the analog of Bethe Ansatz equation for considered case.
The equations (\ref{eq:B}) can be rewritten in the following way:

\bea
(-1)^P\beta_+ \exp\left(-\frac{2\pi ikP}{Q}\right)\prod_{i=1}^Q
	\frac{\Theta_1(z-z_i+{P\over  Q},\tau)}{\Theta_1(z-z_i,\tau)} &+&\nonumber\\
 (-1)^P\beta_- \exp\left(\frac{2\pi ikP}{Q}\right)\prod_{i=1}^Q
	\frac{\Theta_1(z-z_i-{P\over  Q},\tau)}{\Theta_1(z-z_i,\tau)} &- &
\no\\
\alpha_+ \exp\left(\pi i\tau\frac{1+2k}Q-2\pi iz\right)
        \prod_{i=1}^Q \frac{\Theta_1(z-z_i-{\tau\over  Q},\tau)}
	{\Theta_1(z-z_i,\tau)} &-&
\label{eq:Bel}\\
 \alpha_- \exp\left(\pi i\tau\frac{1-2k}Q+2\pi iz\right)
        \prod_{i=1}^Q \frac{\Theta_1(z-z_i+{\tau\over  Q},\tau)}
	{\Theta_1(z-z_i,\tau)} &=& E
\no
\eea

All 4 terms in the sum on the left hand side of (\ref{eq:Bel}) are elliptic
functions, i.e. 2-periodic, in the fundamental domain with vertexes
$(0,\tau,1,1+\tau)$, and each of them has precisely
$Q$ zeroes and $Q$ poles there. Let us suggest that all poles are single.
It is known that such an elliptic function can be written in the form

\be
f(z)=C+\sum_{i=1}^{Q}A_i\zeta(z-z_i),
\label{eq:decom}
\ee
where $\zeta(z)$ is a $\zeta$-function of Weierstrass for the same periodicity
domain. Formula (\ref{eq:decom}) is an analog of decomposition od rational
functions with single poles in terms of functions $\frac{1}{z-z_i}$.

Putting (\ref{eq:decom}) like decomposition for elliptic functions
into (\ref{eq:Bel}) one can verify, that the conditions of cancellation of
the poles precisely are coinciding with the (\ref{eq:Bt}),which are the
equations for $z_i$. Then the energy $E$
is constant function $C$ in this decomposition.

\vspace{1cm}

The authors are acknowledge the Saclay theory division for hospitality,
where this work was started. A.S. thanks P.Wiegmann for discussions.
This work was supported by the Nato linkage grant LG 9303057  and
the grant 211-5291 YPI of the German Bundesministerium f\"ur
	Forschung und Technologie.


\begin{thebibliography}{10}

\bibitem{Zak}
J.Zak.
\newblock {\em Phys. Rev.}, 134:1602, 1964.

\bibitem{Azbel}
M.Azbel.
\newblock {\em Sov.Phys. JETP}, 19:634, 1964.

\bibitem{Wannier}
G.H.Wannier.
\newblock {\em Phys.Status Solidi}, 88:757, 1978.

\bibitem{Hofstadter}
D.R.Hofstadter.
\newblock {\em Phys. Rev.}, B14:2239, 1976.

\bibitem{Wiegmann1}
P.W.Wiegmann and A.V.Zabrodin.
\newblock preprint LPTENS-93/34, September 1993.

\bibitem{Wiegmann2}
P.W.Wiegmann and A.V.Zabrodin.
\newblock Quantum group and magnetic translations. Bethe anzatz for
  Asbel-Hofstadter problem.
\newblock preprint hep-th/94, 1994.

\bibitem{Callan}
D.Freed C.Callan, A.Folce.
\newblock Critical theories of the dissipative Hofstadter model.
\newblock preprint PUPT-1292, CTP-2073 (hep-th/9202085), 1992.

\bibitem{Faddeev}
R.M.Kashaev L.D.Faddeev.
\newblock Generalized Bethe ansatz equations for Hofstadter problem.
\newblock preprint HU-TFT-93-63, December 1993.

\bibitem{Floratos}
Floratos.
\newblock {\em Phys. Lett.}, B233(235), 1989.

\bibitem{Manford}
Manford.
\newblock Tata lectures on theta.
\newblock {\em Progress in Mathematics}, 28, 1983.

\end{thebibliography}
\end{document}